\documentclass[a4paper,10pt]{article}
\usepackage[utf8]{inputenc}
\usepackage{graphicx}
\usepackage{amsmath,amsfonts,amssymb}
\usepackage{hyperref}
\usepackage{color}
\usepackage{csquotes}
\usepackage{authblk}
\usepackage[toc,page]{appendix}

\title{Heavy-quark production with $k_t$-factorization: The importance of the sea-quark distribution}
\author{Benjamin Guiot\footnote{benjamin.guiot@usm.cl}}
\affil[]{\small{Departamento de F\'isica, Universidad T\'ecnica Federico Santa Mar\'ia; Casilla 110-V, Valparaiso, Chile}}
\date{}

\begin{document}

\maketitle

\begin{abstract}
We discuss the fact that $k_t$-factorization calculations for heavy-quark production include only the $gg\rightarrow Q\bar{Q}$ contribution. The cases of
fixed-flavor-number scheme and variable-flavor-number scheme calculations are analyzed separately. For the latter, we show that, similarly to the collinear factorization, the main
contribution is given by the $Qg\rightarrow Qg$ process. In this scheme, calculations including only the $gg$ contribution should show a large discrepancy with the data. 
We show that, if they do not, it is because they include (effectively) a large $K$ factor.
\end{abstract}

\newpage

\tableofcontents

\section{Introduction}
Heavy flavor is an important tool for the study of strong interaction and QCD matter. One reason being that it makes theoretical calculations simpler, in particular because the 
heavy-quark mass allows for the use of perturbation theory. The energy loss of a heavy quark propagating in a medium has been studied extensively, and it is used to determine 
some of the medium properties, like the transport coefficient $\hat{q}$. If using a transport code, the heavy quark mass allows to use equations which are simplified versions of 
the Boltzmann equation. It is a privileged probe for the study of the quark-gluon plasma, since, contrary to light particles, it is generally accepted that it can't be produced 
significantly in the hot medium. Then, the only source of heavy quarks is the hard process, calculable in perturbation theory. From the experimental side, heavy flavors give a clear signal, and experiments like ALICE are able to see the secondary vertex for D mesons.\\

Having a good understanding of heavy-quark production is then of first importance for the phenomenology. In this paper, we concentrate on the $p_t$ distribution of a heavy quark.
More exclusive processes, like quarkonia production, are not considered. We treat the case of collinear factorization in section \ref{hqcoll}, 
where the differences between a fixed-flavor-number scheme and a variable flavor-number scheme are discussed. Some common statements on heavy-quark production will be analyzed, 
and it is reminded \cite{field} that, in a variable-flavor-number scheme\footnote{Probably the most commonly used at LHC energies.}, some of them are wrong. In particular, it
is generally not true that the gluon fusion process gives the main contribution. In the region $p_t>m$, $m$ being the heavy-quark mass, it is in fact given by the 
$Qg\rightarrow Qg$ process.\\

In section \ref{seckt}, we present the usual $k_t$-factorization formula for heavy-quark production. The main goal of this paper is to discuss the fact that $k_t$-factorization calculations
take into account only the $gg\rightarrow Q\bar{Q}$ process. After some remarks, in section \ref{pra}, we analyze separately the cases of fixed-flavor-number scheme
and variable-flavor-number scheme calculations, sections \ref{ktfacffns} and \ref{corrected}. We will see that the situation is similar to the collinear factorization case. When the 
variable-flavor-number scheme is used, the main contribution is given by the $Qg\rightarrow Qg$ process. The $gg$ contribution alone \textit{should not} gives a satisfying description of
the data. If it does, it means that the calculation (effectively) includes a incorrect large $K$ factor, and we will see how it can be ``implemented''. At the end of section \ref{corrected},
numerical calculations using a variable-flavor-number scheme, including flavor excitation processes, are presented.

\section{Heavy flavor production within the collinear factorization \label{hqcoll}}
For hadron-hadron collisions, the collinear factorization formula reads \cite{ElStWe}
\begin{equation}
 \frac{d \sigma}{dx_1dx_2d^2p_t}(P_1,P_2) = \sum_{i,j}f^i(x_1,\mu_f^2)f^j(x_2,\mu_f^2)\hat{\sigma}^{ij}\left(x_1x_2s,p_t,\alpha_s(\mu_R),\frac{Q^2}{\mu_R},\frac{Q^2}{\mu_f}\right), \label{colfac}
\end{equation}
with $s=(P_1+P_2)^2$, $P_k$ the hadron 4-momentum and $x_k$ the longitudinal momentum fraction of the hadron carried by the incoming parton:
\begin{equation}
 x_1=\frac{p_{t}}{\sqrt{s}}e^{y_a}+\frac{p_{t}}{\sqrt{s}}e^{y_b} \hspace{1cm} x_2=\frac{p_{t}}{\sqrt{s}}e^{-y_a}+\frac{p_{t}}{\sqrt{s}}e^{-y_b} \label{xk},
\end{equation}
with $y_a$ and $y_b$ the rapidities of the two outgoing partons and $p_t$ their transverse momentum\footnote{In this study, we will not consider higher order corrections to 
the partonic cross sections, $\hat{\sigma}^{ij}$. Then, the two outgoing partons are back-to-back, both in the partonic COM frame and in the laboratory frame. 
This will not be true anymore when using the $k_t$-factorization, since the transverse momentum, $k_t$, of the incoming partons is taken into account . 
In fact, it is one of the interests of this formalism.}. The hard scale
is denoted by $Q^2$ and is conventionally chosen\footnote{See Ref. \cite{gui} for a detailed discussion on this choice.} to be $p_t^2$. The factorization
and renormalization scales are $\mu_f$ and $\mu_R$, respectively. They are sometimes chosen to be equal, but it is not necessary, and in FONLL calculations \cite{fonll}, they
are varied independently in order to estimate the corresponding uncertainties. The partonic cross sections, $\hat{\sigma}^{ij}$, depend also on the mass of the heavy partons involved in the hard process, and have a perturbative expansion:
\begin{equation}
 \hat{\sigma}=\sum_{k=0}^{\infty}\alpha_s^{n+k}\hat{\sigma}^k,
\end{equation}
with $n$ the power of $\alpha_s$ at leading order. In the case of heavy-quark production, $n=2$. The functions $f^{i}$ are the parton densities.

\subsection{Fixed-flavor-number scheme}\label{ffnss}
\begin{figure}[!h]
\includegraphics[width=12.3cm]{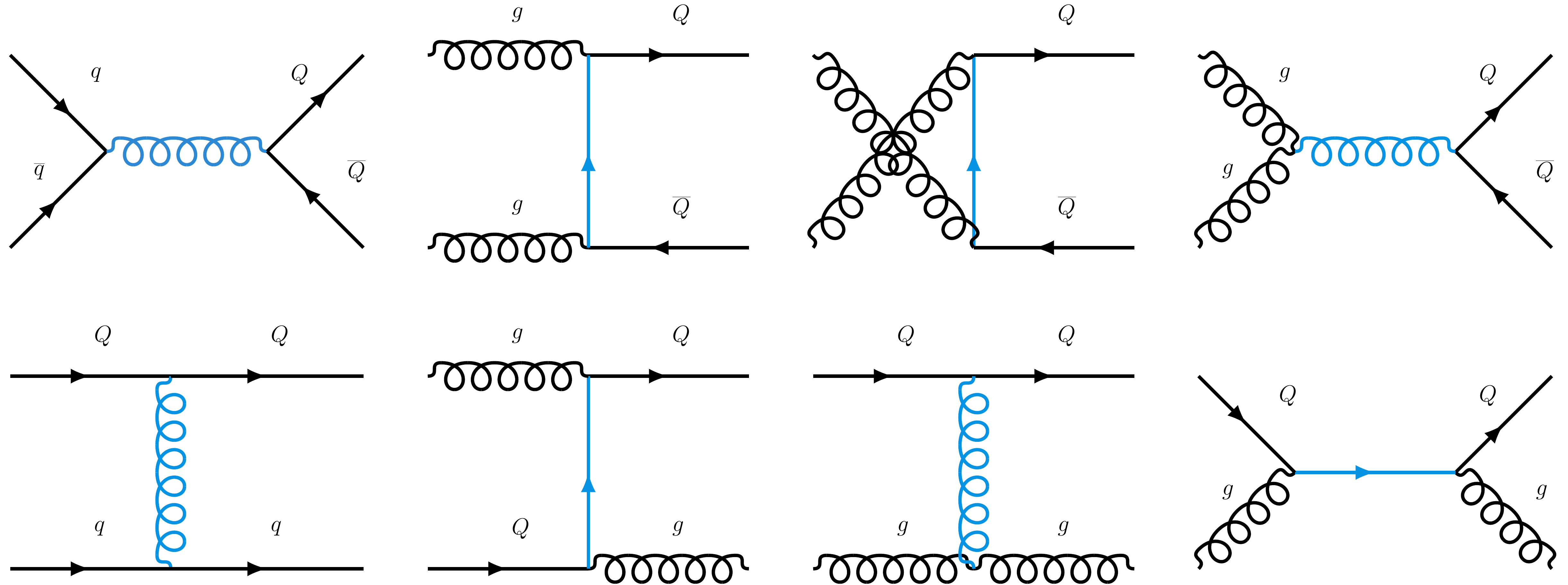}
\caption{Leading order Feynman diagrams with an outgoing heavy quark $Q$ in the VFNS. We neglect contributions implying two incoming heavy quarks like $c\bar{c}\rightarrow c\bar{c}$ or $cb\rightarrow cb$. \label{HQdia}}
\end{figure}

In Eq. (\ref{colfac}), in order to know if the sum over the indices $i$ and $j$ includes the heavy quarks, one has to specify the scheme and the scale $\mu_f$. Historically, the first 
next-to-leading-order (NLO) 
calculations have been done using the fixed-flavor-number scheme (FFNS) \cite{NaDaEl0,NaDaEl,BeKuNe,BeNeMe}. The 3-flavor-scheme assumes that the nucleon is made only of gluons and 
three light quarks, while the 4-flavor-scheme, which can be used for bottom production, also includes the charm quark. Then, LO calculation includes only the flavor creation diagrams shown in the upper row of figure \ref{HQdia}.
It is known \cite{fonll,acot2,KnKrSc1} that this scheme fails at $p_t\gg m_Q$, because of the absence of resummation of the large logarithm $\ln(p_t^2/m_Q^2)$. This issue is solved
by the variable-flavor-number scheme (VFNS), presented in the next section. However, within uncertainties (which are large in this scheme), the NLO FFNS calculations are in good agreement
with data up to quite large $p_t$. For instance, in Ref. \cite{KnKrSc3}, figure 2 (left panel), we observe the agreement of the NLO calculation for bottom production on the full $p_t$ range ([0,25] GeV).\\

The situation is completely different in the case of FFNS LO calculations. Using the same gluon density, a large $K$ factor is necessary to bring agreement with NLO calculations, as discussed in \cite{NaDaEl}.
In this paper, it is shown (figure 12) that at $\sqrt{s}=1.8$ TeV, for the bottom mass, $K=2.5$. The reason for this large factor is not really the ``large'' logarithm, 
 but the fact that NLO contributions open the flavor excitation channels, which have large cross sections.\\

Preparing the discussion on unintegrated PDFs, we note that in \cite{NaDaEl}, the LO calculations without the $K$ factor are below the data\footnote{Below the NLO calculations, and we have
seen that these calculations give a good description of data up to $p_t \backsimeq 5 - 10 m_Q$. See also the NLO results compared to FONLL, figure \ref{bot}.}. But PDFs are
order- and scheme-dependent quantities, and with the choice $g(x,\mu^2)^{LO}=\sqrt{2.5}g(x,\mu^2)^{NLO}$, the LO FFNS calculation for bottom production at the Tevatron
works perfectly fine, since the LO and NLO lines have a similar slope, see \cite{NaDaEl}, figure 12. However, in \cite{NaDaEl} figure 10, we can see that at the same energy, but with
a heavy-quark mass $m_Q=80$ GeV, the $K$ factor is only 1.5, so our new LO FFNS gluon will do a poor description of the (hypothetical) data. It will result in large uncertainties on
the LO FFNS gluon distribution, which is expected since the $\ln(p_t^2/m_Q^2)$ and the flavor excitation cross sections, which partially explain the difference between $m_Q=5$ and $m_Q=80$ GeV, are not at all included.\\

In order to describe heavy-quark production data, using gluon densities with reasonable uncertainties, one should (at least) work 
either with the FFNS at NLO or with the VFNS at LO.

\subsection{Variable-flavor-number scheme}
Even at NLO, calculations using the FFNS do not work quite well at $p_t \gg m_Q$. In this region, it is necessary to resum the large logarithms $\ln(p_t^2/m_Q^2)$. This is achieved
with the VFNS, which includes the heavy-quark density and takes into account the flavor excitation diagrams, even at LO. This scheme is used by the 
GM-VNFS \cite{KnKrSc1,KnKrSc2} and FONLL \cite{fonll} calculations, which include also the resummation of large logarithms due to final state emissions,
using scale-dependent fragmentation functions.\\

The VFNS has several advantages. For $p_t>m_Q$, LO calculations give results comparable to the FFNS NLO calculations, as shown in the next subsection. Compared to the FFNS, the uncertainties due to scale variation
are smaller \cite{KnKrSc3}. Finally, the uncertainties on the gluon densities are also smaller, and going from LO to NLO does not change significantly their value (for $\mu >$ few
GeV, see the two black curves in figure \ref{com}), contrary to the FFNS case.

\subsection{Analysing some common statements on heavy-quark production \label{state}}
In the VFNS, one has to take into account the heavy-quark densities, and we can wonder which process gives the main contribution. For the second part of this paper, about the 
$p_t$ distribution of a heavy quark within $k_t$-factorization, it is useful to analyse first the following common statements in the framework of collinear factorization: 

\begin{enumerate}
 \item At small $x$, the main contribution comes from $gg\rightarrow Q\bar{Q}$.
 \item At small $x$, the gluon distribution is much larger than the charm distribution.
 \item The gluon distribution grows faster than the quark distribution towards small $x$ (see for instance \cite{qcdLev}).
 \item At leading order, one needs a $K$ factor to take into account higher orders corrections ($K>2$).
\end{enumerate}
It is important to understand that there are some implicit statements. For instance, statement 3 is sometimes considered to explain statement 2, and statement 2 is considered to explain 
statement 1. \\

\begin{figure}[!h]
\centering
\includegraphics[width=10cm]{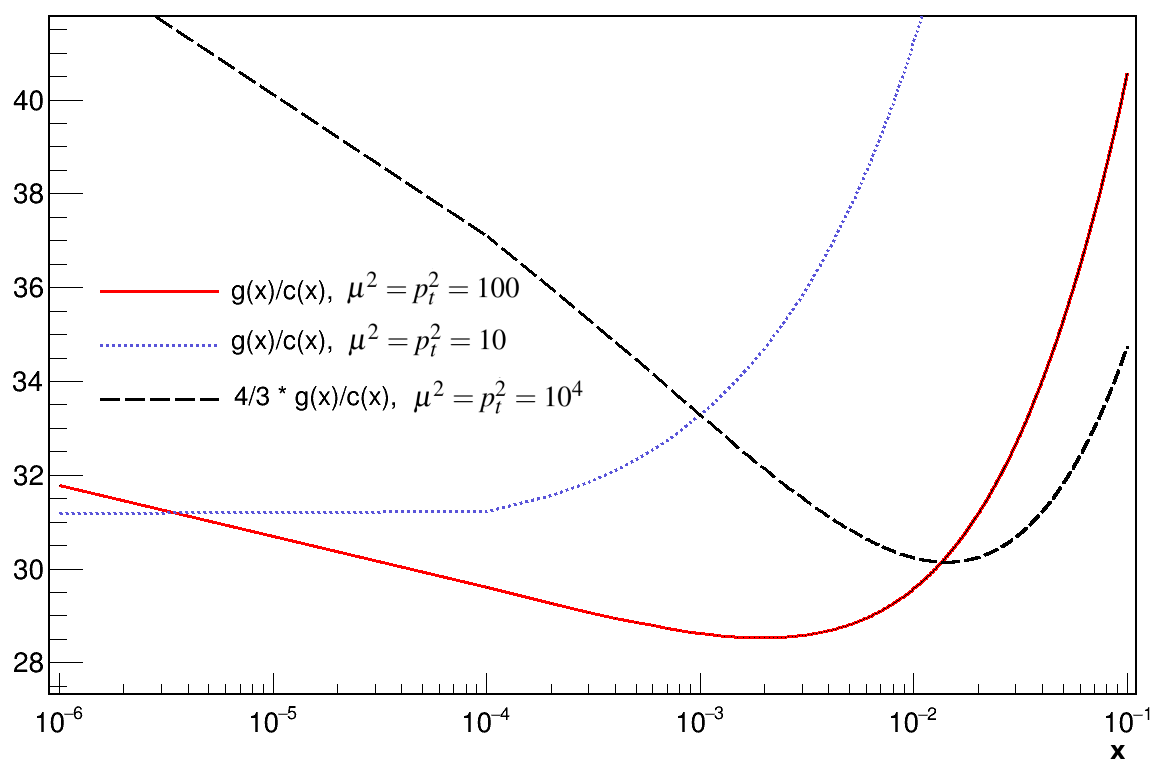}
\caption{Ratio $g(x,\mu^2)/c(x,\mu^2)$ for different $\mu^2$ values, in GeV$^2$. We use the CTEQ14 PDF at NLO \cite{cteq}. It is reminded that for $d\sigma/dp_t$, one usual choice is $\mu_f^2=p_t^2$.\label{rpdf}}
\end{figure}

We will not discuss the correctness of statement 3, since it is a general result of evolution equations \cite{qcdLev} [equation (2.124) and discussion thereafter]. On the contrary, it is wrong to think that statement 3 implies statement 2,
and it is easy to find a counterexample. Consider the two linear functions $f_1(y)=y$, with slope one, and $f_2(y)=2+2y$ with slope two. Even if $f_2$ grows faster than
 $f_1$, the ratio $f_2/f_1$ decreases with $y$.
In figure \ref{rpdf}, we show the ratio $g(x,\mu^2)/c(x,\mu^2)$ for different $\mu^2$ values, obtained with the CTEQ14 PDFs at NLO \cite{cteq} (table CT14n.00.pds).
We see that for $\mu^2=10$ GeV$^2$, the ratio decreases towards small $x$. For $\mu^2=100$ GeV$^2$ the situation is more complicated since it depends on the $x$ range. 
We first have a fast decrease between $x=0.1$ and $x=0.001$. Then, the ratio starts to increase
but very slowly. For $\mu^2=10^4$ GeV$^2$, the increase is faster but this curve corresponds to energies much higher than LHC energies. Generally, one can conclude 
that even if statement 3 is true, it is incorrect to use it in order to justify statement 2. This ratio is large since the beginning, that's all.\\

\begin{figure}[!h]
\centering
\includegraphics[width=6cm]{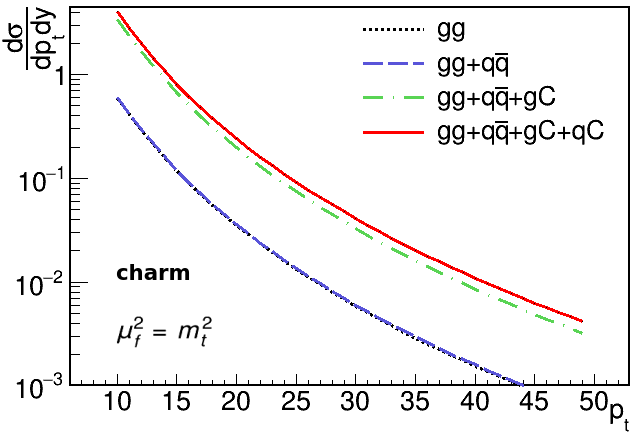}
\includegraphics[width=6cm,height=4.1cm]{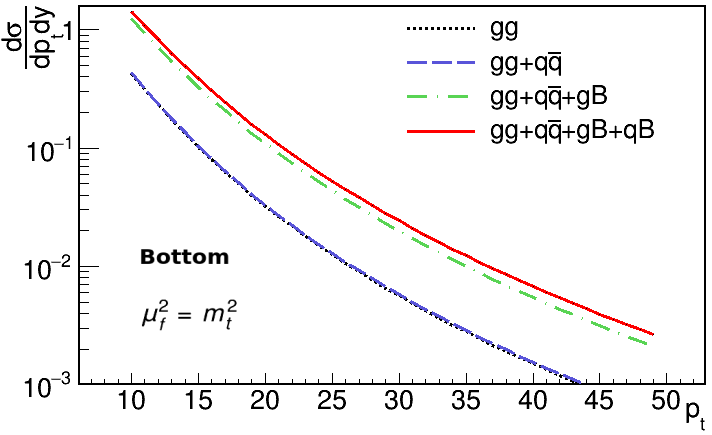}
\caption{Respective contributions for charm and bottom production. These cross sections are given for $-0.5<y<0.5$ and the unit is the $\mu$b. The dotted black and dashed blue
lines are superposed due to the small $q\bar{q}$ contribution.\label{contr}}
\end{figure}

We have seen that statements 3 and 2 are correct, but the former cannot be used to justify the latter (at least at LHC energies). It is also clear that statement 2 alone cannot be used
to justify statement 1. The factorization formula is given by the convolution of parton densities with partonic cross sections, and information on the former is not enough to justify statement 1.
But it is usual to encounter the claim that, due to the high number of gluons, heavy quark production is dominated by the gluon fusion process. Implicitly, it assumes
that the ratio of partonic cross sections $\hat{\sigma}^{ij\rightarrow Q+X}$ is close to one. However, at LHC and RHIC energies, the ratio $\hat{\sigma}^{Qg\rightarrow Qg}/\hat{\sigma}^{gg\rightarrow Q\bar{Q}}$
is large (around 60 at $p_t=10$ GeV, $\sqrt{s}=7$ TeV, and central rapidity). In 2002, Field showed that the main contribution is the
$Qg$ process \cite{field}. A full calculation, based on the implementation of formula (\ref{colfac}), using LO partonic cross sections and CTEQ PDFs \cite{cteq},
is shown in figure \ref{contr}.
We can see that, for the charm quark, the $Qg$ contribution is approximately 4 times larger than the $gg$ contribution, and 3 times larger for the bottom. It was in fact expected, looking at the ratios $g(x,Q^2)/Q(x,Q^2)$ and 
$\hat{\sigma}^{Qg\rightarrow Qg}/\hat{\sigma}^{gg\rightarrow Q\bar{Q}}$, and taking into account the factor 2 for the $Qg$ contribution (the gluon can be provided by hadron 1 
or hadron 2).\\

\begin{figure}[!h]
\centering
\includegraphics[width=8cm]{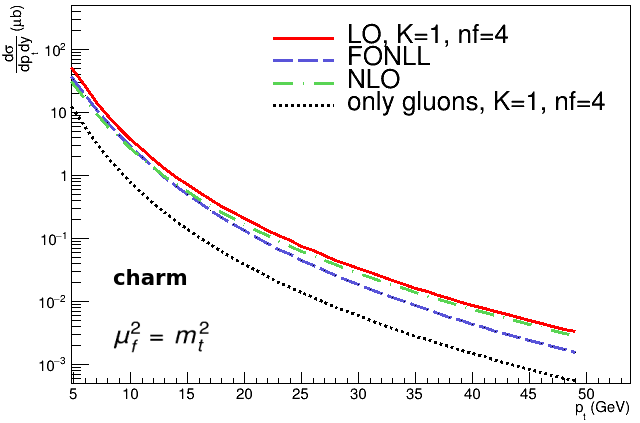}
\includegraphics[width=8cm]{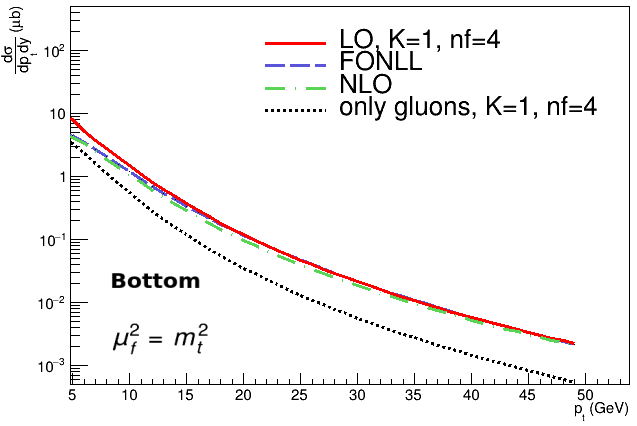}
\caption{Charm and bottom production, obtained with the formula (\ref{colfac}), using LO partonic cross sections, LO CTEQ PDFs \cite{cteq} and $K=1$. This is in good agreement with NLO and FONLL
results, obtained from the web page \cite{foweb}.\label{bot}}
\end{figure}

Finally, let's analyze statement 4. We have already seen that it is true in the FFNS. However, it is not the case in the VFNS, thanks to the heavy-quark density, resuming to all orders large logarithms, and to the flavor excitation diagrams, having large cross
sections. In figure \ref{bot}, we present the results obtained with the collinear factorization 
at LO, for the full contribution (solid red line) and for the $gg$ contribution only (dotted black line). 
Using a factor $K=1$, the full contribution is in very good agreement with NLO\footnote{The NLO result obtained from \cite{foweb} is based on the Nason-Dawson-Ellis calculations
\cite{NaDaEl}.} and FONLL (for the bottom distribution) calculations obtained from \cite{foweb}. In the case of 
charm production, the reason why NLO and our calculations are above the FONLL line is probably because of the absence of resummation of final state emissions. 
In the region $p_t>m_Q$, the gluon fusion contribution completely undershoots the FONLL result. Thinking that the main contribution is given by the $gg$ process, and looking at the LO result (dotted black
line), it is natural to think that at LO, one needs a large $K$ factor. After adding the $Qg$ and $Qq$ contributions, everything is in order. We will see that the situation is similar
within $k_t$-factorization.\\

\section{Standard heavy-quark production within $k_t$-factorization}\label{seckt}
In order to make the comparison with the collinear factorization easier, we rewrite Eq. (\ref{colfac}) as
\begin{equation}
 \frac{d\sigma}{dx_1dx_2d^2p_t}(x_1,x_2,p_t^2,Q^2,\mu^2)=\sum_{ij}f^i(x_1,Q^2;\mu^2)f^j(x_2,Q^2;\mu^2)\hat{\sigma}^{ij}\left(x_1,x_2,p_t^2\right).\label{sigQ2}
\end{equation}
Here, we do not consider the dependence on $\mu_R$ as for simplicity, $\alpha_s$ is taken constant. The factorization scale is now written $\mu$ and, using the freedom on the definition
of parton densities, the logarithms of $Q^2/\mu^2$ have been included in these functions. We keep track of $\mu$ in $f^i$ and $\sigma$ in order to keep in mind that, at finite orders, these quantities do
depend on the factorization scale, giving rise to the factorization scale uncertainty.\\

The $k_t$-factorization (also called high energy factorization or semihard approach) has been developed in parallel in Refs.\cite{ktFac,ktFac2,semihard1,semihard2}, in order to resum
the large logarithms of $\ln(1/x)$ which appear at high energies. For hadron-hadron collisions we have
\begin{multline}
 \frac{d\sigma}{dx_1dx_1d^2p_t}(s,x_1,x_2,p_t^2,\mu^2)=\int^{k_{t,\text{max}}^2}d^2k_{1t}d^2k_{2t} F(x_1,k_{1t}^2;\mu^2)F(x_2,k_{2t}^2;\mu^2) \\
 \times \hat{\sigma}(x_1x_2s,k_{1t}^2,k_{2t}^2,p_t^2), \label{ktfac}
\end{multline}
with the variables $x_1$, $x_2$, $p_t$ and $\mu$ having the same meaning as in Eq. (\ref{sigQ2}). The first difference with the collinear factorization is the use of unintegrated 
parton densities (uPDF), depending on $k_t$, the transverse momentum of the spacelike incoming parton. This additional degree of freedom is integrated out, up to the
kinematical upper bound $k_{t,\text{max}}^2$, discussed in annexe \ref{up}. It is also sometimes necessary to have a specific treatment in the infrared region, see for instance 
Refs. \cite{LiSaZo,ShShSu}. In \cite{ktFac,ktFac2}, the unintegrated gluon density obeys the Balitsky-Fadin-Kuraev-Lipatov (BFKL) equation \cite{bfkl}, while in \cite{semihard2} 
the evolution is given by the nonlinear Balitsky-Kovchegov (BK) equation \cite{bk1,bk2}. In the literature, some studies use equation (\ref{ktfac}) and the terminology $k_t$-factorization, 
but employ uPDFs which do not obey the BFKL or BK equations. In order
to keep the discussion as general as possible, we define the $k_t$-factorization as the convolution of uPDFs with off-shell cross sections, without any condition on the $x$ evolution.
However, we will restrict to the cases where the uPDFs obey
\begin{equation}
 f(x,Q^2;\mu^2)=\int^{Q^2}F(x,k_t^2;\mu^2)d^2k_t, \label{undens}
\end{equation}
[or similar, see for instance Eq. (\ref{inttmd})]. Here, we followed the notation used in Refs. \cite{CaCiHa,CaCiHa2} \footnote{However our function $F(x,k_t^2;\mu^2)$ is related 
to their function by a factor $x$.}. We will also consider the uPDFs which are said to obey approximately Eq. (\ref{undens}).

The second difference with the collinear factorization is the use of off-shell partonic cross sections, which depend on the transverse momenta, $k_{1t}$ and $k_{2t}$, of the incoming 
spacelike partons (with off-shellness $|k_i^2|\simeq k_{it}^2$). Outgoing partons are on shell.

Finally, a third difference, the main purpose of this paper, is the absence of the sum on parton types. The function $F(x,k_t^2;\mu^2)$ corresponds specifically to the unintegrated 
gluon density. This is for instance the case in the following papers \cite{LiSaZo,ShShSu,MaSz,jun02,JuKrLi}. Note that, in few cases, the role of the unintegrated quark density in different processes
has been underlined and studied \cite{CaHa2,CaHa3,HaHeJu}. The off-shell cross section for the process $g^*g^*\rightarrow Q\bar{Q}$, where the stars indicate which parton is 
off shell, can be found in \cite{ktFac2}.\\

In sections \ref{ktfacffns} and \ref{corrected}, we will analyze heavy-quark production within $k_t$-factorization. Before the discussion is in order, 
in section \ref{pra} we present some issues and complications with the use of $k_t$-factorization.

\section{On the practical use of the $k_t$-factorization\label{pra}}
The most problematic part comes from the uPDFs. We first note the existence of various conventions, making the comparison of different papers more complicated. In some cases, it is even
not possible to know which convention has been used, in particular because the relation to the usual PDFs is not given. For instance, if the relation is
\begin{equation}
 f(x,Q^2;\mu^2)=\int^{Q^2}F(x,k_t^2;\mu^2)dk_t^2, 
\end{equation}
then this uPDF is related to the one in Eq. (\ref{undens}) by a factor of $\pi$. Another convention is given below in Eq. (\ref{inttmd}). More important, in view of the next section, 
is the discussion of the uncertainties on these uPDFs. If one finds easily some theoretical explanations or references on how these uPDFs are built, details concerning the 
practical implementation are not always given. It is clear that the implementation of the same uPDF done by different groups can leave to differences in final results. In particular, the uPDFs built from
the usual PDFs, like the KMR \cite{KMR1,KMR2} uPDFs, depend on the choice of the PDF set. We believe that all groups should try to use the same sets of uPDFs\footnote{This
is the case with the collinear factorization where some ``official'' sets are provided by collaborations like CTEQ.}, making the comparison between different studies easier.
For this reason, we think that the TMDlib \cite{tmdlib} is a very good initiative. This is a library of uPDFs, and all the results of this paper will be obtained with uPDFs 
taken from it, when possible.\\

Another issue is the ambiguity of LO calculations, using only unintegrated gluon densities (uGDs). In this case, if not explicitly stated, it is not always possible to know if the calculation
is performed using a FFNS, or using a VFNS with the approximation that the $gg$ process gives the main contribution. We analyse separately these two cases in sections \ref{ktfacffns} and 
\ref{corrected}.\\

\begin{figure}[!h]
\centering
\includegraphics[width=12cm]{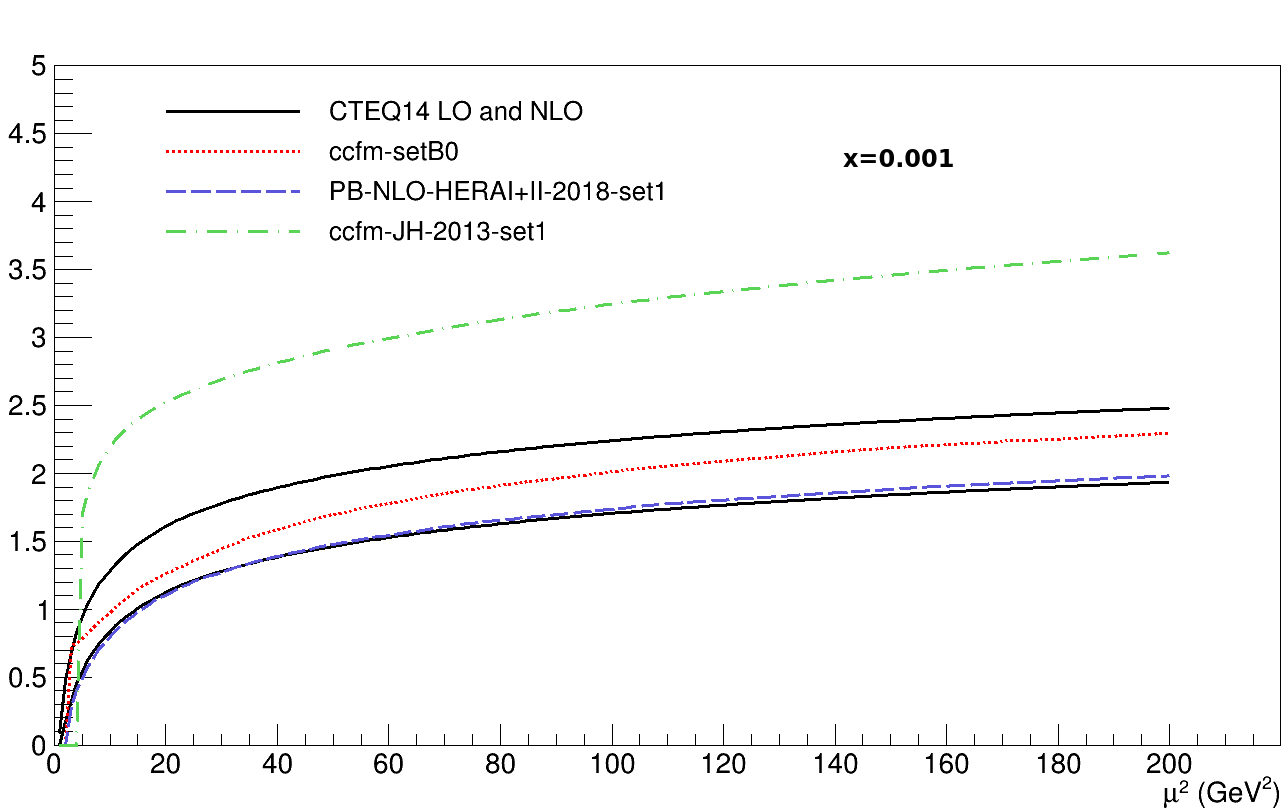}
\caption{CTEQ14 gluon distribution at LO (upper black line) and NLO (lower black line) \cite{cteq}, compared with integrated uPDFs taken from the TMDlib \cite{tmdlib}: ccfm-JH-2013-set1 \cite{ccfmset1}, ccfm-setB0 \cite{JusetB}
and PB-NLO-HERA+II-2018-set1 \cite{pbnlo}. These uPDFs are integrated out following Eq. (\ref{inttmd}).\label{com}}
\end{figure}

Since the parton densities are order and scheme dependent, different choices of schemes can result in quite different uPDFs. 
In figure \ref{com}, we show different uGDs, taken from the TMDlib \cite{tmdlib}, and integrated based on the relation
\begin{equation}
 xg(x,\mu^2;\mu^2)=\int_0^{\mu^2}F(x,k_t^2;\mu^2)dk_t^2. \label{inttmd}
\end{equation}
As a reference, we show the LO and NLO CTEQ14 PDF \cite{cteq} as black lines.
Not surprisingly, the PB-NLO-HERAI+II-2018-set1 (later referred to as PB uPDFs) is in perfect agreement with the NLO CTEQ gluon, since it has been obtained in a VFNS with the
Dokshitzer-Gribov-Lipatov-Altarelli-Parisi (DGLAP)
evolution at NLO. In the opposite, the ccfm-JH-2013-set1 (later referred to as JH uPDFs) is a factor $\sim 2$ above the CTEQ gluons. It is not an issue, and it was in fact expected, since the JH uPDFs have
been determined using a nearly\footnote{The authors have modified the CCFM evolution in order to include valence quarks \cite{ccfmset1}} 0-flavor scheme. Moreover, the CCFM uPDFs are sometimes said
to obey only approximatively to the relation Eq. (\ref{inttmd}).\\

In the next sections, we discuss heavy-quark production within the $k_t$-factorization. We will consider separately  FFNS and VFNS calculations.

\section{$k_t$-factorization with a FFNS \label{ktfacffns}}
Because the Ciafaloni-Catani-Fiorani-Marchesini (CCFM) evolution includes only gluons, calculations using CCFM uGDs is a typical example of FFNS. More precisely, in this case, a 0-flavor scheme. In section \ref{ffnss},
we have seen that LO FFNS gluon densities able to reproduce heavy-quark data at large energies, could be quite larger than NLO FFNS or LO VFNS gluons. It is then not surprising for the
JH uPDFs to be above the CTEQ gluons. After including the NLO contributions in the off-shell cross section, we can expect the CCFM uGDs to be divided by a large factor, 
$K \simeq 2$, similarly to the collinear factorization case\footnote{In particular because, in the limit $k_t\ll p_t$, the off-shell cross sections reduce to the usual ones.}. 
In the next section, it will be explained why the ccfm-setB0 uPDFs (later referred to as B0 uPDFs) are smaller than the JH uPDFs. 

For reasons identical to the collinear factorization case, these calculations suffer from large uncertainties, related to the uGDs and scales variation. The result could be improved,
either by including NLO contributions or by changing to a VFNS\footnote{It possible to modify the CCFM evolution in order to include valence and sea quarks \cite{ccfmset1}.}.
An equivalent statement is that, taking into account the unintegrated heavy-quark density will improve the result.

\section{$k_t$-factorization with a VFNS \label{corrected}}
In the previous section, we discussed that the 0-flavor scheme calculations, which include only the $gg$ contribution, are correct but suffer from large uncertainties. 
Here, the situation is completely different. A VFNS calculation should take into account the flavor excitation processes. Including only $gg\rightarrow Q\bar{Q}$ means that the 
$Qg$ and $Qq$ contributions are considered to be negligible. 

In subsection \ref{kfactor}, we show that this is wrong, and that, similarly to the collinear factorization, the main contribution is
given by $Qg \rightarrow Qg$. Then, in subsections \ref{dispubli} and \ref{kmrsec}, we show that, if available VFNS calculations are in agreement with data, it is because they (effectively)
include a large $K$ factor.

We claim and we will show that, if the $gg$ contribution alone is in agreement with data, the full calculation, including in particular the $Qg$ process, will completely
overshoot the data.

\subsection{Main contribution and large $K$ factor}\label{kfactor}
In subsection \ref{state}, we have seen that in the VFNS at LO, no large $K$ factor is required. We argue that this is also true for $k_t$-factorization. 
Two reasons why no large $K$ factor is required are:
\begin{enumerate}
 \item The arguments given in section \ref{state} are still valid: the flavor excitation cross sections are large and the unintegrated heavy-quark density resums large logarithms. 
 \item Once we take into account all contributions, we obtain a result in good agreement with NLO calculations, see figure \ref{fullkt} and the associated discussion.
\end{enumerate}

\begin{figure}[!h]
\centering
\includegraphics[width=8cm]{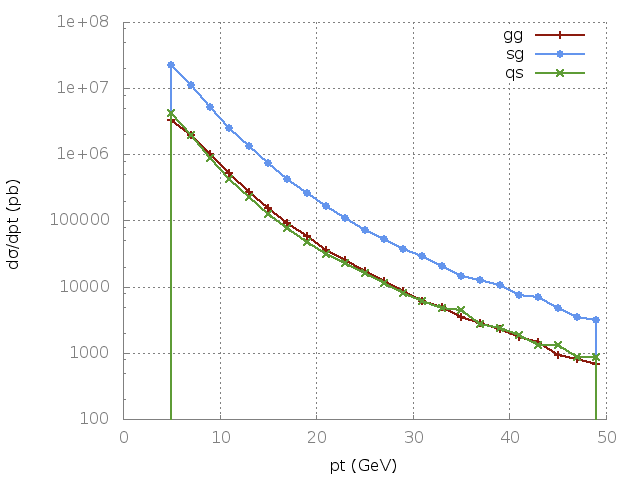}
\includegraphics[width=8cm]{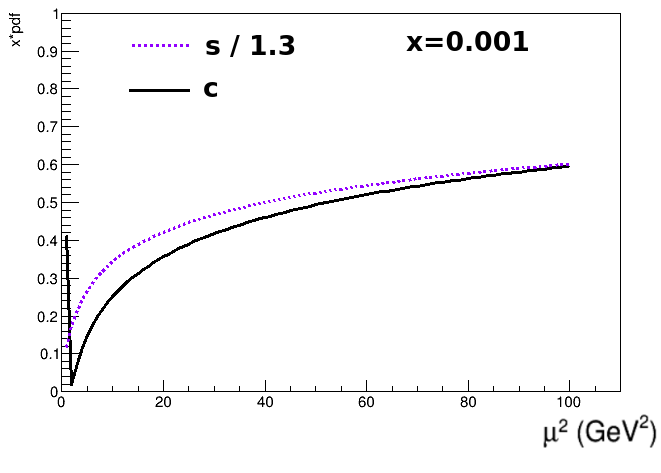}
\caption{Top panel: $gg$, $cg$ and $qc$ contributions to the charm $p_t$ distribution. Instead of the charm uPDF, the strange uPDF divided by 1.3 is used, because \textsc{KaTie} does not 
accept initial heavy states. See the text for more details. Bottom panel: comparison of $c(x,\mu^2)$ and $s(x,\mu^2)/1.3$. \label{sgmain}}
\end{figure}

In figure \ref{sgmain}
are shown the contributions $gg\rightarrow c\bar{c}$, $cg \rightarrow cg$ and $cq \rightarrow cq$ to the charm $p_t$ distribution, obtained with \textsc{KaTie} \cite{katie} and the 
PB uPDFs. The unintegrated gluon distribution from this set has already been presented in figure \ref{com}. The reason why the labels are
$sg$ and $qs$ rather than $cg$ and $qc$ is because heavy quarks in the initial state are not allowed in \textsc{KaTie}. Consequently, we have used $s(x,\mu^2)/1.3$ which is in fact a quite good charm quark,
as shown in the bottom panel of figure \ref{sgmain}. It has been checked that the factor 1.3 does not change to much with $x$. Last detail:
for the calculation of the contributions shown in the top panel of figure \ref{sgmain}, the factorization scale $\mu=(p_t^c+p_t^X)/2$ is used, where $c$ refers to 
the charm and $X$ to the other outgoing particle.
Since we start the calculation for $p_t^c=4$ (and for a technical reason $p_t^X>2$), then $\mu^2\geq 9$ GeV$^2$. Consequently, there is no issue with the behavior of the PDFs at small
$\mu^2$, and it is acceptable to neglect the effect of the charm mass in the matrix elements for $cg\rightarrow cg$ and $cq\rightarrow cq$.

It is clear that the use of $s(x,\mu^2)/1.3$ instead of the true charm distribution includes an uncertainty in our calculations. However, as it can be seen in figure \ref{sgmain},
this uncertainty is of the order of $10\%$, and is irrelevant, since we are discussing the correctness of a large $K$ factor.\\

The result shown in figure \ref{sgmain} is exactly what was expected from the theory. The $cg$ contribution is a factor $\sim 4$ larger
than the $gg$ contribution. This result has been obtained using public codes written by other groups. This is the method used in the case of collinear factorization, where there are few
``official'' PDFs, given by collaborations like CTEQ. The fact that some groups use their unpublished implementation of uPDFs makes it hard or impossible to reproduce their results. In particular,
it is not possible to check if the used uPDFs respect the relation they are supposed to respect, and we will see that it is not always the case.\\

\subsection{Discussion on published results}\label{dispubli}
Calculations using uPDFs determined by the inversion of Eq. (\ref{inttmd}) are an example of VFNS calculations, since the usual PDFs have been obtained in this scheme. This is for instance the case
of the KMR uPDFs, which is treated in detail in the next subsection.
Using uPDFs determined in a VFNS and obeying to Eq. (\ref{inttmd}), it should be impossible to be in agreement with data, taking into account only the $gg$ contribution. 
If the published results do, it is because they (effectively) include a large $K$ factor. It can be done at least in four ways:
\begin{enumerate}
 \item Too large unintegrated gluon distribution, $g(x)\rightarrow \sqrt{K}g(x)$.
 \item Put by hand, in order to take into account supposedly large higher order corrections.
 \item By choosing the factorization scale much higher than the usual choice.
 \item Unreasonably large $k_t$ tail, see subsection \ref{kmrsec}.
\end{enumerate}

By too large uGDs, we mean that, after integrating the uGD with the appropriate formula, the result is a factor $\sqrt{K}$ larger than the appropriate gluon density. By appropriate
gluon density, we mean a gluon density determined in a scheme and to an order identical to the uGD. Due to the lack of information and to the issues discussed in section \ref{pra}, 
it is not always possible to check if the used uPDFs are too large. Ideally, something similar to our figure \ref{com} should be shown, in order to demonstrate that the uPDFs are correctly
normalized. It is done in several papers using CCFM uPDFs, see for instance \cite{ccfmset1}. However, in this case the issue is the lack of gluon densities determined at LO in a 0-flavor scheme\footnote{We
have already mentioned that we expect this gluon density to be quite larger than the CTEQ gluons.}, necessary for a meaningful
comparison.\\

The second possibility consists in adding a factor $K$ by hand, in order to take into account higher order corrections. This factor is discussed 
for instance in Ref. \cite{RyShSh} [equation (21)]. We have already seen that in the VFNS, no large $K$ factor is required. To our knowledge, there is no evidence of the use of such a factor
in recent studies based on the $k_t$-factorization.\\

\begin{figure}[!h]
\centering
\includegraphics[width=10cm]{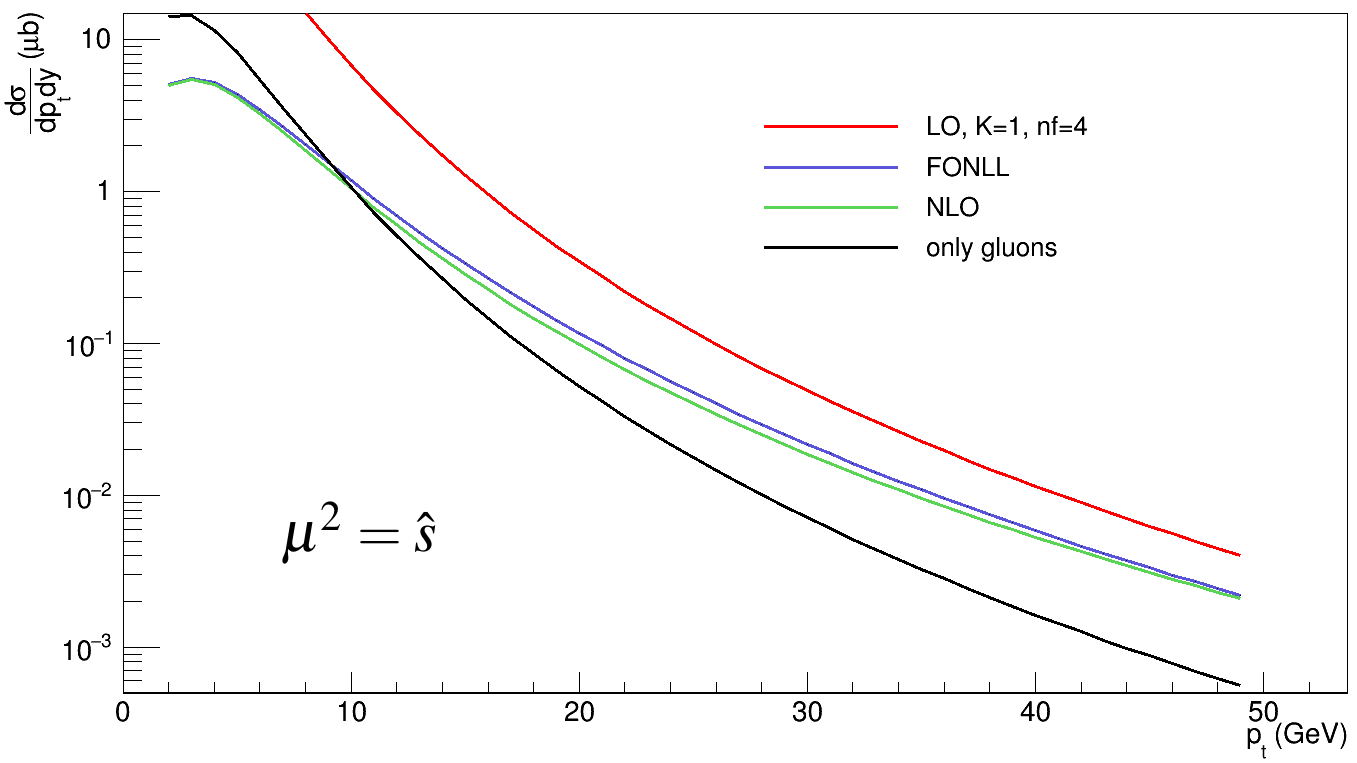}
\caption{Bottom production, obtained with the formula (\ref{colfac}), using LO partonic cross sections, CTEQ PDFs, $K=1$ and the factorization scale $\mu^2=\hat{s}$.\label{bot2}}
\end{figure}

The third possibility consists in using uPDFs with a factorization scale much higher than the usual one, 
$\mu_F^2 \gg \mu^2 \sim p_t^2+m^2$, with $p_t^2$ the transverse momentum of the outgoing parton and $m$ the heavy-quark mass. It is for instance
the case in Ref. \cite{JuKrLi}, where the authors use the B0 uPDFs\footnote{It could look incoherent
to mention CCFM uPDFs in this section dedicated to the VFNS. However, here it is just a numerical matter. The reason why the B0 uGD, 2 times smaller than the JH uGD, gives an 
acceptable result is because they are used with a very large factorization scale, effectively giving a large $K$ factor.}, plotted in figure \ref{inttmd}, with $\mu_F^2=\hat{s}+k_t^2$, ($\hat{s}=x_1x_2s$). In the case of collinear factorization, it is clear that
choosing a much higher factorization scale gives an effective large $K$ factor. Indeed, while physical observables computed to all order do not depend on the factorization scale, in finite order calculations, and in particular at LO, the dependence on $\mu$ can be significant. In figure \ref{bot2}, we show the same calculations as in figure \ref{bot}, now using $\mu^2=\hat{s}$.
Even if this factorization scale is still smaller than the one in Ref. \cite{JuKrLi}, we see that the $gg$ contribution alone is in better agreement with FONLL, while the full contribution
is too high.

Moreover, the factorization scale $\mu_F^2=\hat{s}+k_t^2$ is not appropriate for two reasons. First, it should not be defined as a function of $k_t^2$. Indeed, for uPDFs obeying Eq. (\ref{inttmd}) (or equivalent), it gives the impossible equation:
\begin{equation}
 xg(x,\hat{s}+k_t^2)=\int^{\hat{s}+k_t^2}dk_t^2 F(x,k_t^2;\hat{s}+k_t^2)
\end{equation}
Second, one should avoid defining the factorization scale as a function of $x_1$ and $x_2$, which are integration variables for the cross section $d\sigma/dp_t^2$. In Ref. \cite{CoSoSt}
it is shown that this kind of choice for $\mu$ is dangerous (see the discussion on pages 20--22).\\

Finally, we also want to mention that for D-meson and B-meson production, the branching fraction used for the hadronization is not always indicated.

\subsection{The case of the KMR/MRW parametrization \label{kmrsec}}
\begin{figure}[!h]
\centering
\includegraphics[width=5.3cm,height=5.cm]{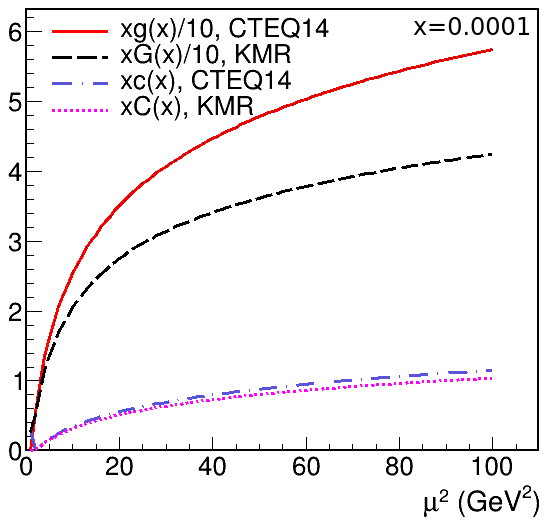}
\includegraphics[width=6.7cm]{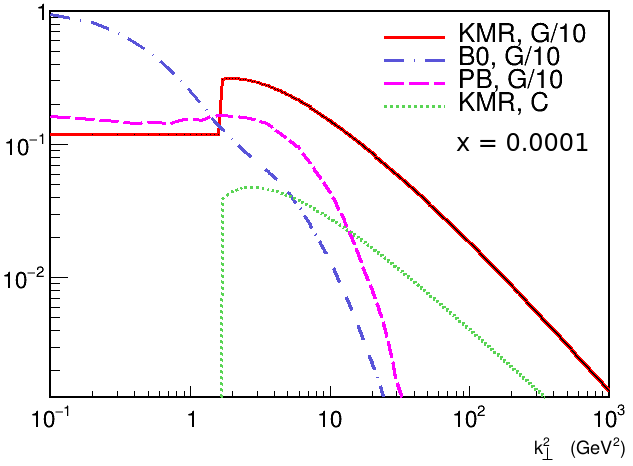}
\caption{Left: based on Eq. (\ref{inttmd}), the KMR uPDFs for charm and gluon are integrated and compared to LO CTEQ14 PDFs. Right: $k_t^2$ distribution of the KRM uPDFs at $\mu^2=10$ GeV$^2$
 and $x=10^{-4}$. For comparison, we also show the result of the ccfm-setB0 and PB-NLO-HERA+II-2018-set1 sets. All the gluon distributions have been normalized by 10. \label{pdfkmr}}
\end{figure}

We have implemented the KMR uPDFs (to be exact, the MRW uPDFs \cite{KMR2}, also used in \cite{MaSz}), using the CTEQ14 LO PDFs. For the gluon, the expression is
\begin{multline}
 F_g(x,k_t^2,\mu^2)=T_g(k_t^2,\mu^2)\frac{\alpha_s(k_t^2)}{2\pi k_t^2}\times \\
 \int_x^1 dz\left[\sum_q P_{gq}(z)\frac{x}{z}q\left(\frac{x}{z},k_t^2\right)+P_{gg}(z)\frac{x}{z}g\left(\frac{x}{z},k_t^2\right)\Theta \left(\frac{\mu}{\mu+k_t}-z\right)\right], \label{FG}
\end{multline}
with $P_{ij}$ the unregularized splitting functions, and $T_g$ the Sudakov form factor:
\begin{equation}
 T_g(k_t^2,\mu^2)=\exp \left(-\int_{k_t^2}^{\mu^2}dq^2 \frac{\alpha_s(q^2)}{2\pi q^2}\left(\int_0^{1-\Delta}dz z P_{gg}(z)+n_f\int_0^1dz P_{qg}(z)\right)\right). \label{tg}
\end{equation}
Note the factor $z$ in front of the $P_{gg}$ splitting function, absent in \cite{KMR1} [equation (3)]. This factor regularizes the divergence of the $P_{gg}$ splitting function at
$x=0$. In order to avoid the divergence at $x=1$, this parametrization uses $z_{\text{max}}=1-\Delta$, with 
\begin{equation}
 \Delta=\frac{k_t}{k_t+\mu}. \label{delta}
\end{equation}
In figure \ref{pdfkmr} we show the result of the implementation at $x=10^{-4}$.
In the left panel, we checked that this parametrization indeed respects the relation (\ref{inttmd}). For gluons, there is a discrepancy of $\sim 25 \%$, due to the introduction of
$\Delta$, not dictated by the DGLAP equation \cite{dgl}. In the right panel, we show
the $k_t^2$ distribution. We have a very similar shape compared to \cite{MaSz}, but our distribution is smaller. It could be due to the fact that we use the CTEQ PDFs, while the 
authors of \cite{MaSz} use the MSTW08 PDFs \cite{mstw}.\\

\begin{figure}[!h]
\centering
\includegraphics[width=8cm]{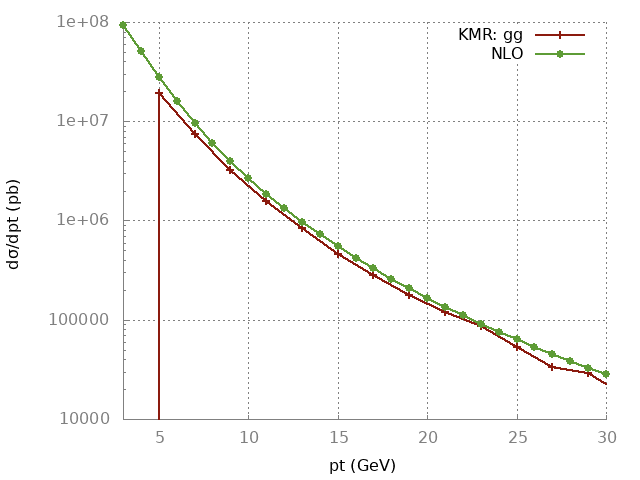}
\caption{NLO calculation for the charm $p_t$ distribution, compared to $gg\rightarrow c\bar{c}$ using KMR gluons.\label{ggkmr}}
\end{figure}

Using our MRW uPDFs and the \textsc{KaTie} event generator, it came as a big surprise to see that the $gg$ contribution alone gives a good description of NLO calculations for the charm 
$p_t$ distribution, as illustrated in Fig. \ref{ggkmr}.
However, as explained before, it is not good news since, after adding the $gQ$ contribution, the result will completely overshoot the NLO line.\\

\begin{figure}[!h]
\centering
\includegraphics[width=5.3cm,height=5.cm]{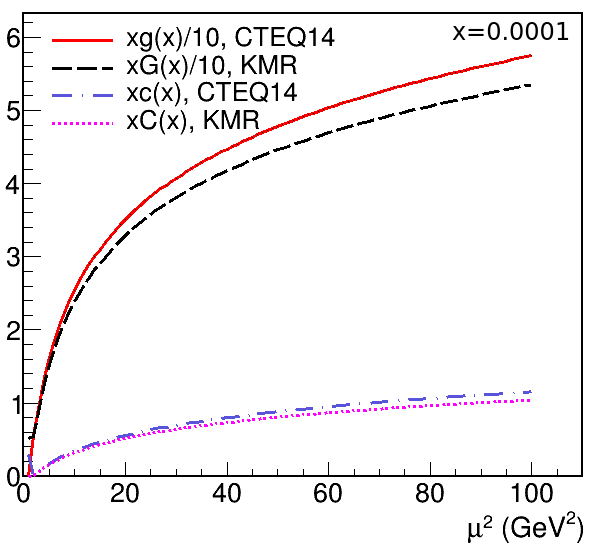}
\includegraphics[width=6.7cm]{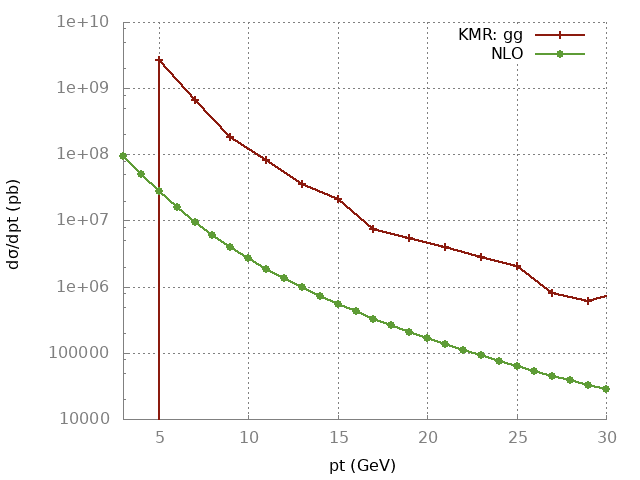}
\caption{Same as figures \ref{pdfkmr} (left) and \ref{ggkmr}, with $z_{\text{max}}=0.99$ ($\Delta=0.01$). \label{kmrzmax}}
\end{figure}

In the following, we will demonstrate that this agreement is accidental. It is related to the definition of $\Delta$, Eq. (\ref{delta}), and to the fact that this specific
implementation of uGD obeys Eq. (\ref{inttmd}) only approximately, as shown in figure \ref{pdfkmr}. At $k_t=\mu$, $z_{\text{max}}=0.5$, and because the parametrization allows 
$k_t>\mu$ (giving a Sudakov form factor larger than one), $z_{\text{max}}$ can even go to zero, an unrealistic value [in Eq. (\ref{tg}), $P_{qg}$ is integrated up to $z=1$]. Using instead
$z_{\text{max}}=0.99$, in Eqs. (\ref{FG}) and (\ref{tg}), we obtained
a much better agreement between the CTEQ and the integrated KMR gluon densities, see Fig. \ref{kmrzmax} (left). However, in this case, the $gg$ contribution overshoots the NLO line
by more than 1 order of magnitude (see Fig. \ref{kmrzmax}, right), showing that the previous agreement was just accidental.

\begin{figure}[!h]
\centering
\includegraphics[width=8cm]{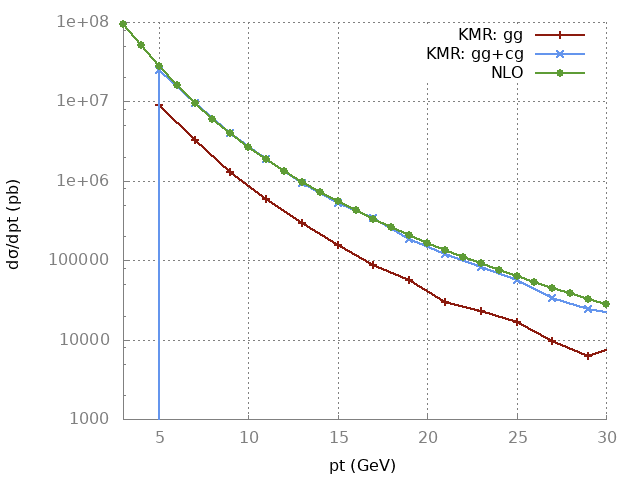}
\caption{NLO calculations for the charm $p_t$ distribution, compared to the $gg$ and $gg$+$cg$ contributions, using our cut KMR uPDFs.\label{kmrcut}}
\end{figure}

The reason for the too large $gg$ contribution, both in figures \ref{kmrzmax} and \ref{ggkmr}, is the large $k_t$ tail of the KMR parametrization. Here, by large we precisely mean
$k_t^2>\mu^2$. This part of the distribution is not constrained by the relation (\ref{inttmd}). While the other uPDFs displayed in figure \ref{pdfkmr} show a very fast decrease at
$k_t^2\gtrsim \mu^2$, the KMR uPDFs decrease slower. To probe the contribution of this large $k_t$ tail, we set the uPDFs equal to zero\footnote{We
didn't choose to set these functions to zero for $k_t^2>\mu^2$ because, compared to the other uPDFs displayed in figure \ref{pdfkmr}, it would have been too rough.} for $k_t^2 > 1.5 \mu^2$. The result obtained with 
these cut KMR uPDFs is shown in figure \ref{kmrcut}.
The $gg$ contribution is now below the NLO line, showing that indeed, the large $k_t$ tail gives an important contribution. Note also the good agreement of the $gg$ + $cg$ calculation.
Clearly, this agreement depends on our choice for the cut KMR uPDFs. What really matters is the confirmation that the $Qg$ process gives the dominant contribution.\\

We believe that the KMR large $k_t$ tail cannot be correct for the following reason. In the KMR paper \cite{KMR1}, the uPDFs are built in two steps. In the first step, inverting Eq. (\ref{inttmd})
and using the DGLAP equation gives
\begin{multline}
 k_t^2F_a(x,k_t^2,\mu^2)=\left. \frac{\partial a(x,\mu^2)}{\partial \ln \mu^2}\right|_{\mu^2=k_t^2}\\
 =\frac{\alpha_s}{2\pi}\sum_{a'}\left[\int_x^{1-\Delta}P_{aa'}(z)a'\left(\frac{x}{z},k_t^2\right)dz-a(x,k_t^2)\int_0^{1-\Delta}P_{a'a}(z)dz\right], \label{revi}
\end{multline}
with $a(x,\mu^2)$ the usual momentum density. The term with a minus sign is referred to as the virtual contribution. In a second step, this virtual contribution disappears, replaced by the
Sudakov form factor, supposed to resum the virtual contribution. It is clear that this Sudakov form factor does not play its role in the region $k_t^2>\mu^2$, 
and while the virtual contribution can be large, leading to a substantial reduction of the first term in the r.h.s. of Eq. (\ref{revi}), the Sudakov form factor (replacing the 
virtual contribution) multiplies this term by a factor larger than 1. This explains the slow decrease of the KMR uPDFs with $k_t^2$.\\

To summarize, with the KMR uPDFs, the agreement of the $gg$ contribution with the NLO result is accidental. The unintegrated gluon does not exactly respect
Eq. (\ref{inttmd}), and, trying to improve the situation by playing with $z_{\text{max}}$ makes the $gg$ contribution overshoot the NLO line. This overestimation is due to the too large $k_t$ tail
of the KMR uPDFs. Consequently, the result displayed in figure \ref{ggkmr} cannot be considered as viable. Having a too large $k_t$ tail could be
seen as another way of implementing a large $K$ factor.

\subsection{Full calculation with \textsc{KaTie} and discussions}\label{subfull}
\begin{figure}[!h]
\centering
\includegraphics[width=10cm]{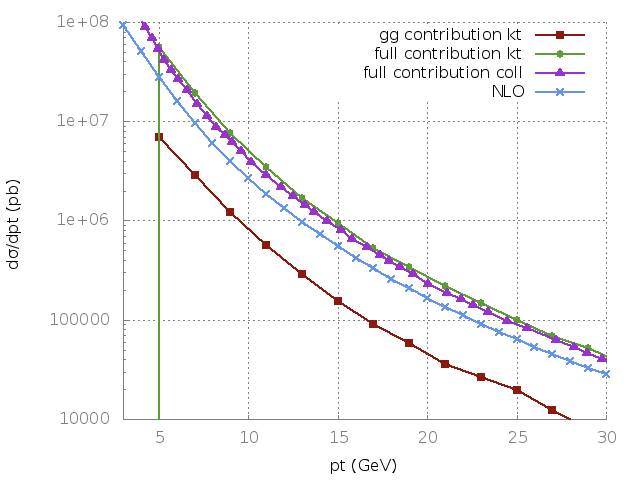}
\caption{Transverse momentum distribution of a charm quark at central rapidity. Results obtained with the \textsc{KaTie} event generator for the $gg$ and the full contributions
are compared with NLO calculations \cite{foweb} and our collinear calculations.\label{fullkt}}
\end{figure}

By employing a VFNS, we have seen that the main contribution to the heavy-quark $p_t$ distribution is the process $Qg \rightarrow Qg$, both within collinear and $k_t$-factorization. Equation (\ref{ktfac}) should be changed for
\begin{multline}
 \frac{d\sigma}{dx_1dx_1d^2p_t}(s,x_1,x_2,p_t^2,\mu^2)=\sum_{ij}\int^{k_{\text{max}}^2}d^2k_{1t}d^2k_{2t} F_i(x_1,k_{1t}^2;\mu^2)F_j(x_2,k_{2t}^2;\mu^2) \\
 \times \hat{\sigma}^{ij}(x_1x_2s,k_{1t}^2,k_{2t}^2,p_t^2), \label{ktfac2}
\end{multline}
where a sum on all parton types has been included. If all contributions are taken into account, no large factor is required. The result for 
the full calculation, using the PB uPDFs and the event generator \textsc{KaTie} is displayed in figure \ref{fullkt}.
The details of the implementation have already been presented in subsection \ref{kfactor}.
While the full calculation is in good agreement with NLO, the $gg$ contribution alone is below the NLO line by a factor $\sim 4$. These calculations also confirm that at medium and large 
$p_t$, collinear and $k_t$-factorizations are numerically close (see \cite{smhard3,gui}).\\

These numerical results confirm the theoretical expectations, and in particular, the importance of the $gQ$ contribution. This conclusion can be probably generalized to 
several phenomenological papers using the $k_t$-factorization. The majority of these papers take into account only the unintegrated gluon density $F_g(x,k_t;\mu)$, while it is likely that 
the unintegrated quark density, $F_q(x,k_t;\mu)$, also plays a non-negligible role in some of these studies. It is then important to systematically explore the effect of the 
quark contribution.\\

Concerning the phenomenology, the $Qg\rightarrow Qg$ process gives kinematical configurations quite different from the $gg\rightarrow Q\bar{Q}$ process.
Including the flavor excitation contributions, as well as spacelike and timelike cascades, is the minimal requirement for a realistic comparison with observables like 
heavy-quarks correlations\footnote{If the observable has been chosen in order to make the contribution of multiple partonic interactions (MPI) negligible. Otherwise a model for MPI
should be implemented for realistic studies.}. Note that, in the framework of models based on collinear 
factorization, the azimuthal correlations between a $b\bar{b}$ pair have been studied in \cite{field}, using the event generators HERWIG, ISAJET and PYTHIA. They observed that 
the toward region ($\Delta \phi \in [0,90]$) is very sensitive to the presence of the flavor excitation and cascades processes.

\section{Conclusion}
We have analyzed the $p_t$ distribution of a heavy quark in the fixed-flavor-number scheme and in the variable-flavor-number scheme. In section \ref{hqcoll}, discussing the case
of collinear factorization, we reminded that
in the FFNS, the NLO contributions give a large $K$ factor, due to the opening of the flavor excitation channel. In the opposite, this is not true in the VFNS, since flavor excitation and 
 the heavy-quark density, resuming to all orders large logarithms of $\ln(p_t^2/m_Q^2)$, are included at leading order.\\

The main goal of this paper was the discussion of the fact that, generally, $k_t$-factorization calculations include only the $gg\rightarrow Q\bar{Q}$ contribution. The conclusion of the discussion
depends on the scheme used. We have seen that in a FFNS, taking into account only the $gg$ contribution is correct, by definition for a 0-flavor scheme. For $n$-flavor schemes
, with $n>0$, it is a good approximation since the $q\bar{q}\rightarrow Q\bar{Q}$ contribution is negligible at small and medium $x$. However, in this scheme, the calculations 
suffer from large uncertainties. Moreover, in the region $p_t\gg m_Q$, NLO FFNS calculations fail and the heavy-quark density has
to be taken into account for accurate predictions. In a VFNS, the unintegrated sea-quark densities should be taken into account, and we have shown that 
the $Qg\rightarrow Qg$ process gives the main contribution, for $p_t>m_Q$. Calculations in agreement with data and taking into account only the $gg$ contribution are incorrect since (1) by definition
of the VFNS, flavor excitation processes should be included and (2) if they were, the obtained result would overshoot data by a large factor (in the region $p_t>m_Q$).\\

In this scheme, if the $gg$ contribution is in agreement with data, it is because the calculation (effectively) includes a large $K$ factor. In subsections \ref{dispubli} and 
\ref{kmrsec}, we discussed how this factor can be implemented. It can be added by hand, or can be obtained by using a too large unintegrated gluon density, a too large factorization
scale or uGDs with a too large $k_t$ tail. We have shown that the latter possibility is the case of the KMR uPDFs.\\ 

In subsection \ref{subfull}, numerical (VFNS) calculations, done with the help of the \textsc{KaTie} event generator and the PB uPDFs have been presented in figure \ref{fullkt}. They show that,
while the $gg$ contribution is far below the NLO line, the full contribution is in fair agreement with NLO calculations. We chose these uPDFs because they are part of the tmdlib library and because
we have been able to check that they do obey the relation they are said to obey, e.g. Eq. (\ref{inttmd}). It is not the case of the KMR parametrization studied in subsection
\ref{kmrsec}, where the uGD shows a disagreement of $\sim 25\%$ with the corresponding gluon density. It is also interesting to note that, for the $p_t$ distribution of a heavy quark, collinear and $k_t$-factorization results are numerically very close, 
in agreement with \cite{smhard3,gui}.\\

Heavy-quark production is probably not an isolated case, and the role of unintegrated quark densities should be systematically studied in papers using the $k_t$-factorization.

\begin{appendices}

\section{Kinematical upper bound\label{up}}
The upper bound for the $k_t^2$ integration is generally not important, since large $k_t$ are suppressed by the unintegrated parton densities, $F(x,k_t^2)$. Sometimes, one finds the condition
\begin{equation}
 (k_1+k_2)^2<s, \label{cond1}
\end{equation}
with  $k_1=[k_0,\overrightarrow{k_t},k_z]$ and $k_2=[k_0,-\overrightarrow{k_t},-k_z]$, in the partonic COM frame. These partons being spacelike, we define
\begin{equation}
 k_1^2=k_2^2=-Q^2. \label{eqspa}
\end{equation}
The condition (\ref{cond1}) is clearly insufficient since here, $k_t$ can go to infinity without violation of this bound or of energy conservation. Indeed, using the approximation $k_t^2=Q^2$ (used in the calculation of off-shell cross sections) we get
\begin{equation}
k_0^2=-Q^2+k_t^2+k_z^2=k_z^2,
\end{equation}
and the energy is finite. Intuitively, the upper bound would be
\begin{equation}
 k_{t,\text{max}}^2=\frac{s}{4}. \label{intuitive}
\end{equation}
In the case of on-shell particles, in order to find the upper bound for $p_t$, one writes an equation for 4-momentum conservation and put $p_z$ to zero. This is how the upper bound $p_t^2<\hat{s}/4$ is found.
Trying to do the same in the case of off-shell particles, we first get
\begin{equation}
 (k_1+k_2)^2=\hat{s}=4k_0^2,
\end{equation}
giving the usual relation $k_0^2=\hat{s}/4$. The second step consists in writing explicitly the relation (\ref{eqspa}):
\begin{equation}
 k_t^2+k_z^2=\frac{\hat{s}}{4}+Q^2.
\end{equation}
In the case of on-shell partons, $Q^2=0$, taking $k_z$ to zero gives $k_{t,\text{max}}^2=\hat{s}/4$. However, in the case of off-shell partons with $k_t^2=Q^2$, the relation becomes
\begin{equation}
 k_z^2=\frac{\hat{s}}{4}.
\end{equation}
We see that it is not possible to obtain the upper bound in this way.\\

\begin{figure}[!h]
\centering
\includegraphics[width=3.5cm]{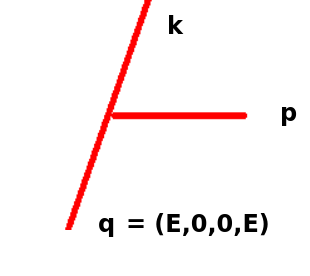}
\caption{p is the 4-momentum of the radiated parton while k is the 4-momentum of the spacelike parton \label{split}. $q$ is the proton (or parton with $x=1$) 4-momentum.}
\end{figure}

We need to find another method for the derivation of the upper bound $k_{t,\text{max}}^2$. First, we want to show that the relation $k_t^2=Q^2$ cannot be correct. We will see that this is an approximation, accurate only in a specific kinematical region. Let's consider the diagram in figure \ref{split}.
Here, we consider the simplified situation where the spacelike parton is generated after the bremsstrahlung from a perfectly collinear on-shell parton with energy 
\begin{equation}
 E=\frac{\sqrt{s}}{2}.
\end{equation}
For the spacelike parton, we choose the parametrization $k=[xE,\overrightarrow{k_t},xE]$, which corresponds to the approximation $k^2=-Q^2=-k_t^2$. But then, 4-momentum conservation
implies that the radiated parton has the 4-momentum
\begin{equation}
p=[(1-x)E,-\overrightarrow{k_t},(1-x)E]. 
\end{equation}
This is not acceptable since it gives $p^2=-k_t^2<0$ and we want the radiated parton to be timelike or on shell. This strange situation is due to the approximation $Q^2=k_t^2$.\\

Relaxing this approximation and using the same diagram, it is in fact possible to derive the true relation between $k_t^2$ and $Q^2$, as well as the upper bounds for these
two quantities. 4-momenta can be written
\begin{eqnarray}
q&=&(E,0,0,E)\\
k&=&(E_k,\overrightarrow{k_t},k_z)\\
p&=&(E_p,-\overrightarrow{k_t},p_z)
\end{eqnarray}
Asking for 4-momentum conservation and choosing the radiated parton on shell gives the following equations:
\begin{eqnarray}
E_p^2&=&k_t^2+p_z^2\\
E_k+E_p&=&E\\
k_z+p_z&=&E
\end{eqnarray}
With these equations, we obtain
\begin{equation}
 E_k+\sqrt{k_t^2+(E-k_z)^2}=E \label{ek}
\end{equation}
\begin{equation}
 E_k^2-k_t^2-k_z^2=2E(E_k-k_z) \label{q2}
\end{equation}
In the lhs of equation (\ref{q2}), one can recognize $k^2=-Q^2$. We then obtain the following expression for the virtuality:
\begin{equation}
 Q^2=2E(xE-E_k), \label{Q2}
\end{equation}
where the definition $k_z:=xE$ has been used. The maximum value is obtained for $E_k=0$:
\begin{equation}
 Q^2_{\text{max}}(x)=2xE^2=\frac{xs}{2}, \label{q2max}
\end{equation}
giving the $x$-dependent upper bound for the virtuality. \\

The relation between $k_t^2$ and $Q^2$ can be obtained using equation (\ref{ek}) and the off-shell condition $E_k^2-k_t^2-k_z^2=-Q^2$, giving
\begin{equation}
 k_t^2=Q^2(1-x)+\frac{Q^4}{s}, \label{ktQ2}
\end{equation}
where we have used the fact that $4E^2=s$ and $k_z=xE$. We see that in the limit $Q^2/s \ll (1-x)$ and $x \ll 1$, this equation reduces to
\begin{equation}
 k_t^2=Q^2. \label{appro}
\end{equation}
Using Eq. (\ref{q2max}), we see that $Q^2/s \ll (1-x)$ is always true at small $x$. Then, the condition $x \ll1$ is enough to ensure the validity of the approximation Eq. (\ref{appro}), showing that at small $x$, it is justified to use this relation when computing the off-shell cross section.\\

As for $Q^2$, the upper bound for $k_t$ is obtained in the case $E_k=0$. Inserting Eq. (\ref{q2max}) in Eq. (\ref{ktQ2}), we obtain
\begin{equation}
 k_{t,\text{max}}^2(x)=\frac{s}{4}\left(2x-x^2\right)
\end{equation}
At small $x$, corresponding to the kinematical region of interest for the $k_t$-factorization, we have
\begin{equation}
 k_{t,\text{max}}^2(x)\simeq \frac{xs}{2}.
\end{equation}
For $x=1$, one has $k_{t,\text{max}}^2=s/4$, which is the $x$-independent intuitive expectation given in Eq. (\ref{intuitive}). This limit corresponds to the simple case where the radiated parton takes all the energy and has no longitudinal momentum, $p=[E,\overrightarrow{E},0]$.
Then the 4-momentum of the spacelike parton is $k=[0,\overrightarrow{E},E]$, showing that the parametrization $[xE,\overrightarrow{k_t},xE]$ can be really incorrect. In this case we have 
$k_t^2=\frac{Q^4}{s}\neq Q^2$. Of course, the probability for an emission with a very large transverse momentum is low, and the region of large $x$ is supposed\footnote{Computing $d\sigma/dp_t$ requires an integration over $x$ which is not restricted to small values. We don't know if the region of, let's say $x>0.01$, gives a negligible contribution.} to be outside of the domain of applicability of the $k_t$-factorization.\\

Finally, we wonder if the upper bound is always irrelevant, if we are only interested by the main contribution. In Ref. \cite{gui}, 
it is shown that doing the integration up to $p_t^2$ (or $m_t^2=p_t^2+m^2$), $p_t$ being the transverse momentum of the outgoing parton, is enough in order to obtain the main 
contribution. If $k_{t,\text{max}}^2 > p_t^2$ or $1 \ll k_{t,\text{max}}^2 < p_t^2$, the upper bound is irrelevant\footnote{The second case is due to the fact that large $k_t$ contributions 
are strongly suppressed by $F(x,k_t^2)$, the unintegrated parton densities.}. Then, we concentrate on the small-$p_t$ and small-$x$ region and wonder when
\begin{equation}
 k_{t,\text{max}}^2(x_{1,2})\simeq \frac{x_{1,2}s}{2}=p_t^2,
\end{equation}
with
\begin{equation}
 x_1=\frac{p_{1,t}}{\sqrt{s}}e^{y_1}+\frac{p_{2,t}}{\sqrt{s}}e^{y_2} \hspace{1cm} x_2=\frac{p_{1,t}}{\sqrt{s}}e^{-y_1}+\frac{p_{2,t}}{\sqrt{s}}e^{-y_2},
\end{equation}
and we choose $p_{1,t}=p_t$, while $p_{2,t}$ is integrated out. Let's take the case of $x_1$. We have
\begin{equation}
 \frac{x_{1}s}{2}=\frac{\sqrt{s}}{2}\left(p_t e^{y_1}+p_{2,t}e^{y_2}\right)=p_t^2.
\end{equation}
The solution is
\begin{equation}
 p_t=\frac{\sqrt{s}}{4}e^{y_1}\left( 1 + \sqrt{1+8\frac{p_{2,t}}{\sqrt{s}}e^{y_2-2y_1}} \right). \label{solpt}
\end{equation}
Let's consider that small-$p_t$ means $p_t=1$ GeV, then the two factors in the r.h.s of Eq. (\ref{solpt}) have to be small. If the term with $p_{2,t}$ is negligible, the condition on $y_1$ is
\begin{equation}
\frac{\sqrt{s}}{2}e^{y_1}=1,
\end{equation}
which corresponds to the value $y_1 \sim -8.16$, for $\sqrt{s}=7$ TeV. If the second term with $p_{2,t}$ is not negligible, the value will be even more negative. For Pb-Pb collision, $\sqrt{s}$ is smaller which gives a smaller value for $y_1$. Then, at the LHC, if one starts to measure particles at $y \sim 7-8$, $k_{t,\text{max}}^2$  will play an important role, while at central rapidities, it does not.\\

It is maybe possible to find a more restrictive $k_{t,\text{max}}^2$, for instance due to angular ordering. In this case, the absolute value of $y$, for which the precise definition 
of the upper bound plays a role, will be smaller. 
\end{appendices}

\section*{Acknowledgement}
We would like to thank Andreas van Hameren for his help with the event generator \textsc{KaTie}. We are also grateful to Klaus Werner, Martin Hentschinski, Francesco Hautmann and Hannes Jung
for their comments on the manuscript. We acknowledge support from Chilean FONDECYT iniciacion grants 11181126. We acknowledge support by the Basal project FB0821.


\end{document}